\documentclass[11pt,a4paper,reqno,twoside]{amsart}
\usepackage{graphicx}    
\usepackage{amssymb}
\usepackage{amsmath}
\usepackage{amsthm}

\newtheorem{theorem}{Theorem}[section]
\newtheorem{lemma}[theorem]{Lemma}
\newtheorem{proposition}[theorem]{Proposition}

\newtheorem{mydef}{Definition}

\newcommand{\ideal}{\mathfrak{a}}
\newcommand{\fielda}{\Bbbk}

\title{An Algebraic Characterization of \\Rainbow Connectivity}

\author[P. Ananth]{Prabhanjan Ananth}
\address{Dept. of Computer Science and Automation, Indian Institute of
Science}
\email{prabhanjan@csa.iisc.ernet.in}
\author[A. Dukkipati]{Ambedkar Dukkipati}
\address{Dept. of Computer Science and Automation, Indian Institute of
Science}
\email{ambedkar@csa.iisc.ernet.in}
\date{}

\begin{document}

\maketitle

\begin{abstract}
The use of algebraic techniques to solve combinatorial problems is studied in this paper. We formulate the rainbow connectivity problem as a system of polynomial equations. We first consider the case of two colors for which the problem is known to be hard and we then extend the approach to the general case. We also give a formulation of the rainbow connectivity problem as an ideal membership problem.  
\end{abstract}

\section{Introduction}
The use of algebraic concepts to solve combinatorial optimization problems has been a fascinating field of study explored by many researchers in theoretical computer science. The combinatorial method introduced by Noga Alon~\cite{alon1999combinatorial} offered a new direction in obtaining structural results in graph theory. Lov\'{a}sz~\cite{lovasz1994stable}, De Loera~\cite{de1995gröbner} and others formulated popular graph problems like vertex coloring, independent set as a system of polynomial equations in such a way that solving the system of equations is equivalent to solving the combinatorial problem. This formulation ensured the fact that the system has a solution if and only if the corresponding instance has a ``yes" answer. 
\par Solving system of polynomial equations is a well studied problem with a wealth of literature on this topic. It is well known that solving system of equations is a notoriously hard problem. De Loera et al.~\cite{de2008hilbert} proposed the NulLA approach (Nullstellensatz Linear Algebra) which used Hilbert's Nullstellensatz to determine the feasibility among a system of equations. This approach was further used to characterize some classes of graphs based on degrees of the Nullstellensatz certificate. 
\par In Section~\ref{background}, we review the basics of encoding of combinatorial problems as systems of polynomial equations. Further, we describe NulLA along with the preliminaries of rainbow connectivity. In Section~\ref{rcidealmember}, we propose a formulation of the rainbow connectivity problem as an ideal membership problem. We then present encodings of the rainbow connectivity problem as a system of polynomial equations in Section~\ref{encodings}. 

\section{Background and Preliminaries}
\label{background}

The encoding of well known combinatorial problems as system of polynomial equations is described in this section. The encoding schemes of the vertex coloring and the independent set problem is presented. Encoding schemes of well known problems like Hamiltonian cycle problem, MAXCUT, SAT and others can be found in~\cite{margulies2008computer}. The term encoding is formally defined as follows:

\begin{mydef}
Given a language $L$, if there exists a polynomial-time algorithm $A$ that takes an input string $I$, and produces as output a system of polynomial equations such that the system has a solution if and only if $I \in L$, then we say that the system of polynomial equations encodes $I$.
\end{mydef} 

It is a necessity that the algorithm that transforms an instance into a system of polynomial equations has a polynomial running time in the size of the instance $I$. Else, the problem can be solved by brute force and trivial equations $0=0$ (``yes" instance) or $1=0$ (``no" instance) can be output. Further since the algorithm runs in polynomial time, the size of the output system of polynomial equations is bounded above by a polynomial in the size of $I$. The encodings of vertex coloring and stable set problems are presented next.

We use the following notation throughout this paper.
Unless otherwise mentioned all the graphs $G=(V,E)$ have the vertex set $V=\{v_1, \ldots ,v_n\}$ and the edge set $E=\{e_1, \ldots ,e_m\}$. The notation $v_{i_1}-v_{i_2}- \cdots -v_{i_s}$ is used to denote a path $\mathcal{P}$ in $G$, where $e_{i_1}=(v_{i_1},v_{i_2}), \ldots ,e_{i_{s-1}}=(v_{i_{s-1}},v_{i_s}) \in E$. The path $\mathcal{P}$ is also denoted by $v_{i_1}-e_{i_1}- \cdots -e_{i_{s-1}}-v_{i_s}$ and $v_{i_1}-\mathcal{P}-v_{i_s}$.

\subsection{$k$-vertex coloring and stable set problem}
The vertex coloring problem is one of the most popular problems in graph theory. The minimum number of colors required to color the vertices of the graph such that no two adjacent vertices get the same color is termed as the vertex coloring problem. We consider the decision version of the vertex coloring problem. The $k$-vertex coloring problem is defined as follows: Given a graph $G$, does there exist a vertex coloring of $G$ with $k$ colors such that no two adjacent vertices get the same color. There are a quite a few encodings known for the $k$-vertex colorability problem. We present one such encoding given by Bayer~\cite{bayer1982division}. The polynomial ring under consideration is $\fielda[x_1, \ldots ,x_n]$.

\begin{theorem}
\label{vertexcolor}
A graph $G=(V,E)$ is $k$-colorable if and only if the following zero-dimensional system of equations has a solution:
\begin{eqnarray*}
x_i^k - 1 &=& 0,\ \forall v_i \in V \\
\sum_{d=0}^{k-1} x_{i}^{k-1-d}x_j^{d} &=& 0,\ \forall (v_i,v_j) \in E
\end{eqnarray*}
\end{theorem}

\noindent \textbf{Proof Idea.} If the graph $G$ is $k$-colorable, then there exists a proper $k$-coloring of graph $G$. Denote these set of $k$ colors by $k^{th}$ roots of unity. Consider a point $p \in \fielda^{n}$ such that $i^{th}$ co-ordinate of $p$ (denoted by $p^{(i)}$) is the same as the color assigned to the vertex $x_i$. The equations corresponding to each vertex (of the form $x_i^k - 1 = 0$) are satisfied at point $p$. The equations corresponding to the edges can be rewritten as $\frac{x_i^{k}-x_j^{k}}{x_i-x_j}=0$. Since $x_i^{k}=x_j^{k}=1$ and $x_i \neq x_j$, even the edge equation is satisfied at $p$. 
\par Assume that the system of equations have a solution $p$. It can be seen that $p$ cannot have more than $k$ distinct co-ordinates. We color the vertices of the graph $G$ as follows: color the vertex $v_i$ with the value $p^{(i)}$. It can be shown that if the system is satisfied then in the edge equations, $x_i$ and $x_j$ need to take different values. In other words, if $(v_i,v_j)$ is an edge then $p^{(i)}$ and $p^{(j)}$ are different. Hence, the vertex coloring of $G$ is a proper coloring.\\ 

\noindent A stable set (independent set) in a graph is a subset of vertices such that no two vertices in the subset are adjacent. The stable set problem is defined as the problem of finding the maximum stable set in the graph. The cardinality of the largest stable set in the graph is termed as the independence number of $G$. The encoding of the decision version of the stable set problem is presented. The decision version of the stable set problem deals with determining whether a graph $G$ has a stable set of size at least $k$. The following result is due to Lov\'{a}sz~\cite{lovasz1994stable}.

\begin{lemma}
A graph $G=(V,E)$ has an independent set of size $\geq k$ if and only if the following zero-dimensional system of equations has a solution
\begin{eqnarray*}
x_i^{2} - x_i &=& 0,\ \forall i \in V\\
x_ix_j &=& 0,\ \forall \{i,j\} \in E\\
\sum_{i=1}^{n} x_i - k &=& 0\ .
\end{eqnarray*}
The number of solutions equals the number of distinct independent sets of size $k$. 
\end{lemma} 

\noindent The proof of the above result can be found in~\cite{margulies2008computer}.

\subsection{NulLA algorithm}
\label{nulla}
De Loera et al.~\cite{de2008hilbert} proposed the Nullstellensatz Linear Algebra Algorithm (NulLA) which is an approach to ascertain whether the system has a solution or not. Their method relies on the one of the most important theorems in algebraic geometry, namely the Hilbert Nullstellensatz. The Hilbert Nullstellensatz theorem states that the variety of an ideal is empty over an algebraically closed field iff the element 1 belongs to the ideal. More formally,
\begin{theorem}
Let $\ideal$ be a proper ideal of $\fielda[x_1, \ldots ,x_n]$. If $\fielda$ is algebraically closed, then there exists $(a_1, \ldots ,a_n) \in \fielda^{n}$ such that $f(a_1, \ldots ,a_n)=0$ for all $f \in \ideal$.  
\end{theorem}

Thus, to determine whether a system of equations $f_1=0, \ldots ,f_s=0$ has a solution or not is the same as determining whether there exists polynomials $h_i$ where $i \in \{1, \ldots ,s\}$ such that $\sum_{i=1}^{s} h_if_i=1$. A result by Koll{\'a}r~\cite{kollar1988sharp} shows that the degree of the coefficient polynomials $h_i$ can be bounded above by $\mathrm{deg}\{3,d\}^{n}$ where $n$ is the number of indeterminates. Hence, each $h_i$ can be expressed as a sum of monomials of degree at most $\mathrm{deg}\{3,d\}^{n}$, with unknown coefficients. By expanding the summation $\sum_{i=1}^{s} h_if_i$, a system of linear equations is obtained with the unknown coefficients being the variables. Solving this system of linear equations will yield us the polynomials $h_i$ such that $\sum_{i=1}^{s} h_if_i=1$. The equation $\sum_{i=1}^{s} h_if_i=1$ is known as Nullstellensatz certificate and is said to be of degree $d$ if $\mathrm{max}_{1 \leq i \leq s} \{ \mathrm{deg}(h_i)\}=d$. There have been efforts to determine the bounds on the degree of the Nullstellensatz certificate which in turn has an impact on the running time of NulLA algorithm. The description of the NulLA algorithm can be found in \cite{margulies2008computer}. The running time of the algorithm depends on the degree bounds on the polynomials in the Nullstellensatz certificate. It was shown in \cite{brownawell1987bounds} that if $f_1=0, \ldots ,f_s=0$ is an infeasible system of equations then there exists polynomials $h_1, \ldots ,h_s$ such that $\sum_{i=1}^{s} h_i f_i =1$ and $\mathrm{deg}(h_i) \leq n(d-1)$ where $d=\mathrm{max}\{\mathrm{deg}(f_i)\}$. Thus with this bound, the running time of the above algorithm in the worst case is exponential in $n(d-1)$. Even though this is still far being practical, for some special cases of polynomial systems this approach seems to be promising. More specifically this proved to be beneficial for the system of polynomial equations arising from combinatorial optimization problems~\cite{margulies2008computer}. Also using NulLA, polynomial-time procedures were designed to solve the combinatorial problems for some special class of graphs~\cite{loera2009expressing}. 

\subsection{Rainbow connectivity}
Consider an edge colored graph $G$. A rainbow path is a path consisting of distinctly colored edges. The graph $G$ is said to be rainbow connected if between every two vertices there exists a rainbow path. The least number of colors required to edge color the graph $G$ such that $G$ is rainbow connected is called the rainbow connection number of the graph, denoted by $rc(G)$. The problem of determining $rc(G)$ for a graph $G$ is termed as the rainbow connectivity problem. The corresponding decision version, termed as the $k$-rainbow connectivity problem is defined as follows: Given a graph $G$, decide whether $rc(G) \leq k$. The $k$-rainbow connectivity problem is NP-complete even for the case $k=2$.

\section{Rainbow connectivity as an ideal membership problem}
\label{rcidealmember}
Combinatorial optimization problems like vertex coloring~\cite{alon1997note,de1995gröbner} were formulated as a membership problem in polynomial ideals. The general approach is to associate a polynomial to each graph and then consider an ideal which contains all and only those graph polynomials that have some property (for example, chromatic number of the corresponding graph is less than or equal to $k$). To test whether the graph has a required property, we just need to check whether the corresponding graph polynomial belongs to the ideal. In this section, we describe a procedure of solving the $k$-rainbow connectivity problem by formulating it as an ideal membership problem. By this, we mean that a solution to the ideal membership problem yields a solution to the $k$-rainbow connectivity problem. We restrict our attention to the case when $k=2$. 
\par  In order to formulate the $2$-rainbow connectivity problem as a membership problem, we first consider an ideal $I_{m,3} \subset \mathbb{Q}[x_1, \ldots ,x_m]$. Then the problem of deciding whether the given graph $G$ can be rainbow connected with $2$ colors or not is reduced to the problem of deciding whether a polynomial $f_G$ belongs to the ideal $I_{m,3}$ or not. The ideal $I_{m,3}$ is defined as the ideal vanishing on $V_{m,3}$, where $V_{m,3}$ is defined as the set of all points which have at most 2 distinct coordinates. More formally, $V_{m,3} \in \mathbb{Q}^{m}$ is the union of $S(m,2)$ (Stirling number of the second kind) linear subspaces of dimension 2. The following theorem was proved by De Loera~\cite{de1995gröbner}:

\begin{theorem}
The set of polynomials $\mathcal{G}_{m,3}=\{ \prod_{1 < r < s \leq k} (x_{i_r}- x_{i_s})\ |\ 1 < i_1 < i_2 < i_3 < m\}$ is a universal Gr\"obner basis\footnote{A set of generators of an ideal is said to be a universal Gr\"obner basis if it is a Gr\"obner basis with respect to every term order.} for the ideal $I_{m,3}$.
\end{theorem}         

\noindent We now associate a polynomial $f_G$ to each graph $G$ such that $f_G$ belongs to the ideal $I_{m,3}$ if and only if the rainbow connection number of the graph $G$ is at least 3. Assume that the diameter of $G$ is at most 2, because if not we have $rc(G) \geq 3$. We first define the path polynomials for every pair of vertices $(v_i,v_j) \in V \times V$ as follows: If $v_i$ and $v_j$ are adjacent then $P_{i,j}=1$, else
$$P_{i,j}=\sum_{e_a,e_b \in E:\ v_i-e_a-e_b-v_j \in G} (x_a - x_b)^2\ .$$ 
The polynomial $f_G$ is nothing but the product of path polynomials between any pair of vertices. Formally, $f_G$ is defined as follows:
$$f_G=\prod_{v_i,v_j \in V;\ i < j} P_{i,j}$$

\noindent Note that $f_G$ can be computed in polynomial time.
\begin{theorem}
The polynomial $f_G \in I_{m,3}$ if and only if $rc(G) \geq 3$. 
\end{theorem}
\begin{proof}
To prove the theorem, it is enough to show that $\forall p \in V_{m,3}$, $f_G(p) = 0$ if and only if rainbow connection number of $G$ is at least 3. Assume that the rainbow connection number of $G$ is at most 2. This means that there exists an edge coloring of the graph with two colors such that the graph is 2-rainbow connected. We can visualize this coloring of edges as a tuple  $(c_1,  \ldots ,c_m)$ where $c_i \in \mathbb{Q}$ and the edge $e_i$ is given the color $c_i$. It can be seen that the point $p=(c_1,  \ldots ,c_m)$ belongs to $V_{m,3}$. We claim that $f_G(p) \neq 0$. For that, we show that $P_{i,j}(p) \neq 0$ for all $(v_i,v_j) \in V \times V$. 
Assume that $v_i$ and $v_j$ are not adjacent (this is because $P_{i,j}(p) \neq 0$ for adjacent pair of vertices $(v_i,v_j)$). Since $G$ is rainbow connected, there is a rainbow path from $v_i$ to $v_j$ and let $e_a,e_b$ be the two edges in this path. Correspondingly, $(c_a-c_b)$ is non-zero and hence $(c_a-c_b)^2$ is positive. This implies that $P_{i,j}(p) \neq 0$ for every pair of vertices $(v_i,v_j)$. Hence, $f_G(p)$ is non-zero. 
\par Assume that $f_G(p) \neq 0$ for some $p=(c_1, \ldots ,c_m) \in V_{m,3}$. Using $p$, we color the edges of the graph $G$ with two colors such that $G$ is rainbow connected. Assume without loss of generality that $b$ and $r$ are the only two values taken by the entries in $p$. Color the edges of $G$ as follows: If $c_i=b$ then color the edge $e_i$ with blue else color the edge $e_i$ with red. Since, $f_G(p) \neq 0$ we have $P_{i,j}(p) \neq 0$ for all $i,j \in \{1, \ldots ,n\}$. Consider a non adjacent pair of vertices $(v_i,v_j)$. This implies that there exists $a$ and $b$ such that $(x_a-x_b)^2$ is in the support of $P_{i,j}$ and $(c_a-c_b)^2$ is non-zero. Correspondingly, the path from $v_i$ to $v_j$ containing the edges $e_a$ and $e_b$ is a rainbow path since $e_a$ and $e_b$ are colored distinctly. Thus, $G$ is rainbow connected which implies that $rc(G) \leq 2$.  
\end{proof}

\noindent We now have a procedure to decide whether the rainbow connection of a graph is at most 2 or not. 
\begin{itemize}
\item [1.] Given a graph $G$, find its corresponding polynomial $f_G$. 
\item [2.] Divide $f_G$ by $\mathcal{G}_{m,2}$.
\item [3.] If the division algorithm gives a non-zero remainder then the rainbow connection number of the graph is at most 2 else $rc(G) \geq 3$ . 
\end{itemize}

\section{Encoding of rainbow connectivity}
\label{encodings}

\noindent Consider the polynomial ring $\mathbb{F}_2[x_1, \ldots ,x_m]$. As before, assume that the diameter of $G$ is at most 2. We present an encoding of the 2-rainbow connectivity problem as a system of polynomial equations $S$ defined as follows: 
$$\prod_{e_a,e_b \in E: v_i-e_a-e_b-v_j \in G} \left( x_a + x_b + 1 \right) =\ 0;\ \forall i,j \in \{1, \ldots ,n\}, i < j,\ (v_i,v_j) \notin E$$


If all pairs of vertices are adjacent (as in the case of clique), we have the trivial system $0=0$.

\begin{proposition}
The rainbow connection number of $G$ is at most 2 if and only if $S$ has a solution.
\end{proposition}
\begin{proof}
Let $p=(c_1, \ldots ,c_m) \in \mathbb{F}_{2}^m$ be a solution to $S$. Consider the edge coloring $\chi:E \rightarrow \{\mbox{blue},\mbox{red}\}$ defined as follows: $\chi(e_i)=\mbox{blue}$ if $c_i=1$ else $\chi(e_i)=$ red. Now, consider a pair of vertices $(v_i,v_j) \notin E$. Since the equation corresponding to $(i,j)$ is satisfied at $p$, there exists $a$ and $b$ such that $e_a$ and $e_b$ are edges in the path from $v_i$ to $v_j$ and $c_a+c_b+1=0$. This implies that $c_a$ and $c_b$ have different values and hence the edges $e_a$ and $e_b$ are colored differently. In other words there is a rainbow path between $v_i$ and $v_j$. Since, this is true for any pair of vertices, the graph $G$ is rainbow connected. 
\par Assume that $rc(G) \leq 2$. Then, let $\chi:E \rightarrow \{\mbox{blue,red}\}$ be an edge coloring of $G$ such that $G$ is rainbow connected. Let $p=(c_1, \ldots ,c_m)$ be a point in $\mathbb{F}_2^{m}$ such that $c_i=1$ if $\chi(e_i)=$blue else $c_i=0$. The claim is that $p$ is a solution for the system of polynomial equations $S$. Consider a pair of non adjacent vertices $(v_i,v_j)$ in $G$. Since $G$ is rainbow connected there exists a rainbow path from $v_i$ to $v_j$. Let $e_a$ and $e_b$ be the edges on this path. Since these two edges have distinct colors, correspondingly the expression $c_a+c_b+1$ has the value zero. In other words, the point $p$ satisfies the equation corresponding to $i,j$. Since this is true for any pair of vertices the point $p$ satisfies $S$. 
\end{proof}

\noindent \textit{Example.} Consider a graph $G_n=(V,E)$ such that $V=\{a,v_1, \ldots ,v_n\}$ and $E=\{(a,v_i)\ |\ i \in \{1, \ldots ,n\} \}$. It can be easily seen that the rainbow connection number of the graph $G_n$, for $n \geq 3$, is at least 3. We show this by using the system of equations denoted by $S$ as follows. The system of equations $S$ for $G_n$, for $n \geq 3$, is given by:
$$e_i+e_j+1=0,\ \ \ \ \ \ \ \ \forall i,j \in \{1, \ldots ,n\}, i < j\ .$$ 
Since $(e_1+e_2+1)+(e_2+e_3+1)+(e_1+e_3+1)=1$, we have the fact that 1 belongs to the ideal $\ideal=\langle e_i+e_j+1\ :\ \forall i,j \in \{1, \ldots ,n\}, i < j\rangle$. By Hilbert Nullstellensatz, this means that the solution set of $\ideal$ is empty which further implies that the system of equations $S$ defined for $G_n$, for $n \geq 3$, has no solution. From the above proposition, we have the result that the rainbow connection number of $G_n$ is at least 3.\\   


\noindent We now generalize the encoding for the 2-rainbow connectivity problem to the $k$-rainbow connectivity problem. We will only consider graphs of diameter at most $k$. This encoding is similar to the one described for the $k$-vertex coloring problem. The polynomial ring under consideration is $\mathbb{C}[x_1, \ldots ,x_m]$.

\begin{theorem}
\label{rc}
The rainbow connection number of a graph $G=(V,E)$ is $\leq k$ if and only if the following zero-dimensional system of equations has a solution:
\begin{eqnarray*}
x_i^k - 1 &=& 0,\ \forall e_i \in E \\
\prod_{v_i-\mathcal{P}-v_j} \left( \sum_{e_a,e_b \in \mathcal{P}} \left( \sum_{d=0}^{k-1} x_{a}^{k-1-d}x_b^{d} \right)^{k} \right) &=& 0,\ \forall (v_i,v_j) \notin E
\end{eqnarray*}
\end{theorem}

\begin{proof}
The proof is similar to that of Theorem~\ref{vertexcolor}. Assume that the system of polynomial equations has a solution $p$. We color the edges of the graph as follows: Color the edge $e_i$ with $p^{(i)}$ ($i^{th}$ coordinate of $p$). Consider a pair of non adjacent vertices $(v_i,v_j) \in V \times V$. Corresponding to this pair, there is an equation in the system which is satisfied at $p$. This implies that for some path $\mathcal{P}$ between $v_i$ and $v_j$, the polynomial  $\sum_{e_a,e_b \in \mathcal{P}} \left( \sum_{d=0}^{k-1} x_{a}^{k-1-d}x_b^{d} \right)^k$ vanishes to zero at point $p$. This further implies that $ \left( \sum_{d=0}^{k-1} x_{a}^{k-1-d}x_b^{d} \right)^k$ is zero for any pair of edges $e_a,e_b$ on the path $\mathcal{P}$. This can happen only when $p^{(a)}$ is different from $p^{(b)}$. Correspondingly any two edges $e_a$ and $e_b$ on the path $\mathcal{P}$ are assigned different colors. Thus the path $\mathcal{P}$ between vertices $v_i$ and $v_j$ is a rainbow path. This is true for all pairs of vertices and hence the graph is rainbow connected. Since the point $p$ has at most $k$ distinct coordinates (this is because $p$ satisfies equations of the form $x_i^k - 1 = 0$), we have the rainbow connection number of $G$ to be at most $k$. 

Let the rainbow connection number of graph $G$ be at most $k$. We find a point $p$ belonging to the solution set of the given system of polynomial equations. As in the case of proof of Theorem~\ref{vertexcolor}, denote the $k$ colors by $k^{th}$ roots of unity. Let $p \in \mathbb{C}^{m}$ such that the entry $p^{(i)}$ of $p$ is equal to the color assigned to the edge $e_i$. The set of equations $x_i^k-1=0$ are satisfied at $p$. Consider a pair of vertices $(v_i,v_j) \notin E$ in graph $G$. Since graph $G$ is $k$-rainbow connected, there is a rainbow path $\mathcal{P}$ between $v_i$ and $v_j$. Consider any two edges $e_a$ and $e_b$ on the path $\mathcal{P}$. Since $e_a$ and $e_b$ are colored differently, the indeterminates $x_a$ and $x_b$ are given different values. This further implies that the expression $\sum_{d=0}^{k-1} x_{a}^{k-1-d}x_b^{d}$ is zero. Thus, for a rainbow path $\mathcal{P}$ between $v_i$ and $v_j$, the summation $\sum_{e_a,e_b \in \mathcal{P}} \left( \sum_{d=0}^{k-1} x_{a}^{k-1-d}x_b^{d} \right)^{k}$ is zero and hence, the equation corresponding to the pair of vertices $(v_i,v_j)$ is satisfied at point $p$. Since this is true for any pair of vertices, the point $p$ satisfies the given system of polynomial equations.   
\end{proof}

The above given formulation of the $k$-rainbow connectivity problem, for any $k$, as a system of polynomial equations is not a valid encoding since the encoding procedure does not run in time polynomial in $n$. However, if $k$ is a constant then we have a polynomial time algorithm to exhaust all the paths of length at most $k$ between every pair of vertices. Using this, we can transform the graph instance into a system of polynomial equations in time polynomial in $n$. Hence if $k$ is a constant, Theorem~\ref{rc} gives a valid polynomial time encoding of the $k$-rainbow connectivity problem. 

\section{Conclusion}
In this paper, we reviewed methods to solve graph theoretic problems algebraically. One of the most popular being formulation of the combinatorial problems as a system of polynomial equations. Using this formulation, an approach to determine the infeasibility of the system of polynomial equations, namely NulLA, is described. We solve the rainbow connectivity problem in two ways. We formulate the problem as a system of polynomial equations and using NulLA this will give a solution to our original problem. We also formulate the problem as an ideal membership problem such that determination of whether the graph can be colored with some number of colors is equivalent to determining whether a specific polynomial belongs to a given ideal or not.
\par An interesting future direction might be to analyze the special cases for which the rainbow connectivity problem is tractable using the above characterization (the rainbow connectivity problem is NP-hard for the general case). In order to achieve this, it would be interesting to get some bounds on the degree of the Nullstellensatz certificate for the polynomial system corresponding to the rainbow connectivity problem.  

\bibliographystyle{plain}
\bibliography{prabhanjan}

\end{document}